# The effect of varying atmospheric pressure upon habitability and biosignatures of Earth-like planets

Engin Keles[a*], John Lee Grenfell[b], Mareike Godolt[a],

Barbara Stracke[b], Heike Rauer[a,b]

[a] Zentrum für Astronomie und Astrophysik, Technische Universität Berlin, Hardenbergstraße 36, 10623 Berlin, Germany

[b] Institut für Planetenforschung, Deutsches Zentrum für Luft- und Raumfahrt, Rutherfordstraße 2, 12489 Berlin, Germany

*Corresponding author, email: keen08@physik.fu-berlin.de

# Effect of Pressure on Earth-like planets

## Abstract

Understanding the possible climatic conditions on rocky extrasolar planets, and thereby their potential habitability, is one of the major subjects of exoplanet research. Determining how the climate, as well as potential atmospheric biosignatures, change under different conditions is a key aspect when studying Earth-like exoplanets. One important property is the atmospheric mass hence pressure and its influence on the climatic conditions. Therefore, the aim of the present study is to understand the influence of atmospheric mass on climate, hence habitability, and the spectral appearance of planets with Earth-like, that is, $N_2$-$O_2$ dominated, atmospheres orbiting the Sun at 1 Astronomical Unit. This work utilizes a 1D coupled, cloud-free, climate-photochemical atmospheric column model; varies atmospheric surface pressure from 0.5 bar to 30 bar; and investigates temperature and key species profiles, as well as emission and brightness temperature spectra in a range between 2μm - 20μm. Increasing the surface pressure up to 4 bar leads to an increase in the surface temperature due to increased greenhouse warming. Above this point, Rayleigh scattering dominates and the surface temperature decreases, reaching surface temperatures below 273K (approximately at ~34 bar surface pressure). For ozone, nitrous oxide, water, methane, and carbon dioxide, the spectral response either increases with surface temperature or pressure depending on the species. Masking effects occur, for example, for the bands of the biosignatures ozone and nitrous oxide by carbon dioxide, which could be visible in low carbon dioxide atmospheres.

**Effect of Pressure on Earth-like planets**

**1. Introduction**

Many factors influence planetary habitability such as the host star and planetary orbital distance, the planetary atmosphere, water reservoir, planetary mass, and possibly plate tectonics. In this work, the influence of different atmospheric masses and, hence, surface pressure upon habitability is investigated. This is motivated by the fact that the atmospheric mass of exoplanets is not known. The surface pressure of Venus, for example, is ~92 times higher than that of Earth (although both planets have similar mass and radius) indicating a different atmospheric development. The surface pressure of terrestrial planets depends on several factors, for example on outgassing processes (e.g., Elkins- Tanton & Seager (2008)), escape processes (e.g., Lammer et al. (2013)), impacts (e.g., de Niem et al. (2012)), or surface deposition. Modeling planetary atmospheres helps to understand the relevant processes for habitability as well as processes that influence the appearance of atmospheric spectral signatures in atmospheres. To date, more than 3000 confirmed extrasolar planets (www.exoplanet.eu, Nov. 2016) are known. A key aim of exoplanetary research is to detect potentially habitable planets and investigate whether there are any biosignatures in their atmospheres. A possibility for detection of these indications for life is to analyze emitted or transmitted light from the atmosphere of an exoplanet. This detected light can reveal indications of biotic species in the atmosphere and under some circumstances information with regard to atmospheric properties such as surface pressure and temperature (e.g., Benneke & Seager, 2012; Kaltenegger & Selsis, 2007; Rugheimer et al., 2013). So called spectral biosignatures (e.g., from ozone, nitrous oxide, oxygen) that may indicate life are of main interest. Nitrous oxide on Earth is mainly ( ≥ 99 %) produced by (de)nitrifying bacteria, and in this sense it is an excellent indication for life. Although methane on Earth is mainly produced by methanogenic bacteria, it is not a direct indication for life because it can also be produced abiotically, for example, by volcanism. It is therefore termed a bioindicator. Another biosignature

**Effect of Pressure on Earth-like planets**

is ozone (noting that it can also be produced abiotically; see, e.g., Luger and Barnes, 2015), which protects the lower atmosphere from UV-radiation and leads to the stratospheric temperature maximum in the mid-stratosphere. While water vapor (hereafter referred to as water) is not a biosignature, a planet with a liquid water reservoir on the surface will show water vapor in its atmosphere. Furthermore, water is one of the major greenhouse gases that determine planetary climate. Detecting only one of the above biosignatures may not be a sufficient indication of life. Rather, several atmospheric compounds may need to be observed simultaneously as suggested by Selsis et al. (2002). For example, the simultaneous detection of ozone and the life-related compounds carbon dioxide and water could constitute evidence of life (Selsis et al., 2002). Also the combined detection of huge amounts of oxygen and methane, that is, oxidizing and reducing species, has been regarded as possible evidence for life (Sagan et al., 1993).

Atmospheric modeling with increased surface pressure for terrestrial planets has been done, for example, for the exoplanet candidate Gliese 581d to investigate habitable conditions, assuming the potential for a carbon dioxide rich atmosphere (Wordsworth et al., 2010; von Paris et al., 2010; Kaltenegger et al., 2011). Such modeling has been attempted for early Mars as well, assuming high carbon dioxide content, with the intent to investigate the effect upon atmospheric temperature (Kasting, 1993; von Paris et al., 2013; Ramirez et al. (2014a); Wordsworth et al., 2012, Wordsworth et al., 2017). Different pressures of Earth-like atmospheres have been investigated, for example, with a focus on the runaway greenhouse effect (Kasting, 1988), probing atmospheric pressure by modeling spectroscopic measurements of $O_2$-$O_2$ dimers (Misra et al., 2014) or for dry, rocky planets at the inner edge of the habitable zone (Zsom et al., 2013). For planetary habitability, it has been shown that increasing the atmospheric mass may increase the surface temperature via an increased greenhouse effect (Kopparapu et al. (2014);Wolf &

**Effect of Pressure on Earth-like planets**

Toon (2014)). However, a larger atmospheric mass may also lead to enhanced cooling due to increased Rayleigh scattering. The cooling effect induced by Rayleigh scattering was shown, for example, by Kasting et al. (1993), Zsom et al. (2013), and von Paris et al. (2013). Vladilo et al. (2013, 2015) analyzed the habitability of an Earth-like planet (with $N_2$-$O_2$ dominated atmosphere) with enhanced pressure but without considering Rayleigh scattering.

In this study, we used a 1D atmospheric column model to investigate the effect of varying pressure on Earth-like $N_2$-$O_2$ dominated atmospheres of planets with the same biomass development and planetary properties as Earth, with consideration for the interaction between greenhouse warming and Rayleigh scattering. We intended to accomplish two main goals when initiating this work. The first goal was to analyze physical and chemical interactions on varying atmospheric pressure and assess their impact on habitability and biosignatures. The second main goal was to show the spectral response of the simulated atmospheres and investigate the behavior of key chemical species for different pressure scenarios, for example, with regard to visibility, masking, and shape change of the spectral bands. The approach in the present study is similar to those of Grenfell et al. (2007), Rauer et al. (2011), and Segura et al. (2003).

The organization of this paper is as follows: Section 2 gives a model description. Section 3 discusses the simulated scenarios and shows the model input parameters. Section 4 investigates the results and Section 5 provides a discussion. Section 6 gives a summary & conclusion.

## 2. Model Description

We make use of a steady-state, cloud-free, 1D coupled photochemical and climate model, which calculates globally and diurnally averaged parameters (see, e.g., Kasting et al., 1980; Segura et al., 2003; Grenfell et al., 2007; Rauer et al., 2011 for a detailed description). To calculate the

**Effect of Pressure on Earth-like planets**

spectral response, the line-by-line model SQuIRRL (Schwarschild Quadrature InfraRed Radiation Line-by-line) (Schreier und Schimpf (2001)) is used. The 1D model consists of two parts, the chemistry and the climate modules, which are coupled. In short, the converged temperature-, pressure- and water-profiles calculated by the climate model are passed to the chemistry model, and then converged radiative gas concentrations from the chemistry model are passed back to the climate model and so on, until the temperature, radiative flux, and species concentration are converged. The main start and boundary parameters for the climate and chemistry model are shown in the Appendix Table A2 . The surface albedo is adjusted due to the fact that clouds are not considered in the simulation so that the surface temperature reaches ~288K for a modern Earth simulation.

2.1 Climate Module

The climate module calculates (using start values taken from the U.S. standard atmosphere 1976 and assuming constant gravity through the atmosphere) the pressure-, water- and temperature-profiles. To calculate the pressure-profile, hydrostatic equilibrium is assumed. The surface pressure depends on the density of $N_2$, $O_2$, Ar, $CO_2$, and $H_2O$ on the surface. From the pressure profile, the altitude is calculated and divided into 52 layers. The temperature-profile is calculated numerically via an iterative process by solving the radiative transport equation downwards from the top model layer (TOA) at $6.6 \times 10^{-5}$ bar for each layer and applying convective adjustment. For convective layers the wet adiabatic lapse rate formulation (Kasting (1988)) is used to recalculate the temperature. If heat transport is given by radiative transport, the radiative lapse rate is calculated by solving the radiative transfer equation separately for the thermal (von Paris et al. (2015)) and the solar radiation (Toon et al. (1989)). The surface layer, however, is always set to

**Effect of Pressure on Earth-like planets**

be convective. Table 3 (appendix) shows the considered species and effects in the model for the solar and thermal radiative transfer scheme. The solar radiation is covered by 38 spectral bands (~2200 cm$^{-1}$ – 42087 cm$^{-1}$) with the absorption of carbon dioxide, methane, oxygen, ozone, and water considered. The angular integration is accomplished by using a 2-stream approximation for the radiative fluxes. At the TOA, the incoming thermal flux is set to zero. The climate module uses an improved thermal transfer scheme as described by von Paris et al. (2015). The thermal radiation transport scheme uses the correlated k-method (similar to that of Mlawer et al. (1997)) to calculate molecular absorption in 25 spectral bands (20 cm$^{-1}$ – 50 000 cm$^{-1}$) and is adapted for a temperature range of 100K-400K. In all bands, $CO_2$ and $H_2O$ are considered, as well as $CH_4$ and $O_3$ in their main spectral ranges. By using k-distributions up to 1.5 bar surface pressure, the absorption coefficients of water, carbon dioxide, methane, and ozone are calculated. Beyond this range, the absorption coefficients are extrapolated. The scheme features an update in the $CO_2$ continuum absorption; recalculation of heat capacities; recalculation of Rayleigh scattering cross-sections of CO, $CH_4$, $H_2$, and He; new cross-sections for optical depth calculations; and $H_2O$ and $CO_2$ self and foreign continuum absorption. The water amount up to the tropopause depends on the atmospheric temperatures and is calculated from a fixed relative humidity distribution. Above the tropopause, the water amount is calculated by the chemistry module.

2.2 Chemistry Module

The chemistry module solves the 1D continuity equation to reach equilibrium by using more than 200 chemical reactions and 55 atmospheric species. The atmospheric species molecular nitrogen, molecular oxygen, and carbon dioxide are fixed as isoprofiles. The module features 64 atmospheric layers from the surface up to 6.6 x 10$^{-5}$ bar. The module calculates the chemical

**Effect of Pressure on Earth-like planets**

reactions and the species' concentrations in each layer by using reaction rates for uni, bi-, and termolecular reactions. Chemical emission, dry and wet deposition, and Eddy-diffusion through the atmosphere are considered. Sources of NO via lightning, as well as $H_2S$ and $SO_2$ via volcanism, are included in the troposphere. A downward effusive flux of CO and O is included in the top layer and surface deposition and emission of, for example, $CH_4$, $H_2$, CO, $N_2O$, and $CH_3Cl$. The continuity equation is solved by using the backwards Euler method.

2.3 Spectral model

To calculate the spectral response, the line-by-line model SQuIRRL (Schreier und Schimpf (2001); Schreier et al. (2014)) was used assuming local thermal equilibrium and using temperature, pressure, and species' concentration profiles calculated with the coupled atmospheric column model. The spectra are calculated by using the Hitran Database 2008 for the wavelength range between 2 μm - 20 μm, assuming a pencil beam looking downward from the top of atmosphere at $6.6 \times 10^{-5}$ bar (for a detailed description see Hedelt et al., 2013; Rauer et al., 2011; Schreier & Böttger, 2003).

**3. Scenarios**

Nine scenarios were performed (see Table 1), the surface pressure varying from 0.5 bar to 30 bar (0.5 bar, 1 bar, 2 bar, 3 bar, 4 bar, 5 bar, 10 bar, 20 bar, 30 bar). For all scenarios, species' concentration profiles, temperature profiles, radiative flux profiles, and molecular emission spectra are shown. For higher surface pressure (of approximately ~34 bar), the surface was no longer habitable ($T_0 < 273K$), although 3D model simulations show that habitable regions are

**Effect of Pressure on Earth-like planets**

possible (Godolt et al. (2016)). For detailed analysis, brightness temperature spectra for the 1bar, 4 bar, and 30 bar simulations are shown with the presence and absence of the investigated species. The modern Earth control run (1 bar) reproduces modern Earth conditions in a similar manner as calculated by Grenfell et al. (2011). These scenarios are chosen because 1 bar represents the modern Earth case, the 4-bar scenario shows the highest increase in surface temperature, and the 30-bar scenario is the highest pressure scenario that enables global-mean habitable conditions. The surface albedo is set to 0.2155, which produces a modern Earth surface temperature of ~288 K orbiting the Sun at 1 AU. Boundary and start parameters (Table 4 in the appendix) similar to those of the modern Earth run are used (similar to Grenfell et al., 2007; Rauer et al., 2011).

**4. Results**

Table 2 shows the resulting surface temperatures in Kelvin and column densities in Dobson units for all scenarios. A detailed discussion of the results is given in sections 4.1 and 4.2.

4.1 Climatic response to surface pressure variations

Figure 1a shows the surface temperatures and 1b temperature profiles for the different surface pressure scenarios. The surface temperature first increases (up to 4 bar surface pressure) up to a value of 295 K and then decreases for higher pressure scenarios. The temperature profiles for higher and lower surface pressures exhibit a similar vertical form as that of modern Earth. The tropospheric adiabatic lapse rate (dT/dP) first increases in the scenarios 1 to 6 but then decreases for the subsequent scenarios on increasing surface pressure. The lapse rate changes are related to

**Effect of Pressure on Earth-like planets**

changes in water vmr and temperature. The tropopause occurs at around 0.1 bar as expected (see Robinson & Catling 2014). The stratospheric profiles are similar. This is an indication that ozone heating does not change significantly (*4.2.1*). The temperature at the stratopause increases for higher surface pressure. For the mesosphere, the pressure-temperature profile is similar for scenarios with surface pressure higher than 1 bar.

Vladilo et al. (2013) simulated Earth-like atmospheres with an EBM (Energy Balance Model) neglecting the scattering effect. In their work, increased surface pressures lead to increasing surface temperatures. Figure 2 shows a comparison between our results and those of Vladilo et al. (2013). To reproduce the results of Vladilo et al. (2013) (red line), we performed high pressure simulations neglecting both Rayleigh scattering and chemistry, that is, with fixed species volume mixing ratios similar to those of Vladilo et al. (2013). For a better comparison to their results without Rayleigh scattering, we increased the surface albedo to 0.292. Because of this, we reached a surface temperature of ~288 K for 1 bar surface pressure.

In these scenarios, the surface temperature increases with surface pressure, and the resulting temperatures are broadly comparable. When considering Rayleigh scattering, the surface temperature decreases (green line) for the higher pressure simulations. Vladilo et al. (2015) found a similar behavior in surface temperatures for pressures above 3 bar using a more sophisticated ESTM (Earth-like planet surface temperature model) to investigate the habitability of Kepler-62e (see their Figure 4). A similar effect, where increased pressure leads to tropospheric cooling induced by Rayleigh scattering for an Earth-like carbon dioxide increased atmosphere has been shown, for example, by Ramirez et al. (2014b) and by Kasting et al. (1993) for an Earth-like planet at the outer edge of the habitable zone. This has also been shown by von Paris et al. (2013) for early Mars and by Zsom et al. (2013) for desert worlds at the inner edge of the habitable zone.

**Effect of Pressure on Earth-like planets**

Figure 3a shows the difference of the net upward propagating thermal radiation at the surface and the TOA converted to a temperature difference by the Stefan-Boltzmann law. This is a measure of the greenhouse effect of the atmosphere, which increases up to scenario 6 and decreases for the other scenarios. Figure 3b shows the planetary albedo, which steadily increases with surface pressure leading to a cooling of the atmosphere for the higher pressure scenarios. The increase (scenarios 1 to 5) and decrease (scenarios 6 to 10) in surface temperature result from the net effect of these two processes. The values of the 1-bar run for the greenhouse effect and the planetary albedo deviate from those measured for the present Earth, which is mainly because we apply a cloud-free model.

4.2 Effect of surface pressure on chemical profiles

The increase in atmospheric mass leads to a change in radiative flux, as well as in temperature and pressure, which affects the chemical reactions and thereby the atmospheric composition. The following text is an analysis of the response of the species ozone, nitrous oxide, water, and methane.

*4.2.1 Ozone*

The ozone profiles (Fig. 4a) for varying surface pressure simulations show a similar vertical form as the 1-bar scenario. One can see the "ozone layer" that peaks at 0.04 bar and the "smog" ozone in the troposphere. Although for scenario 9, for example, the surface pressure is thirty times larger than that for modern Earth conditions, the ozone column density only changes by less than a factor of two (Table 2). Ozone amounts are subject to a stabilizing ozone-, UV-feedback. There

**Effect of Pressure on Earth-like planets**

is a trade-off situation between the availability of UV-flux and molecular oxygen both of which are needed to produce ozone, which changes with altitude. The production of ozone near the surface is catalyzed by volatile organic carbon species (e.g. Sillman (1999)) as well as $NO_X$. The production of "smog" ozone increases for higher pressure scenarios and leads to higher ozone column densities. But this does not, however, lead to a significantly stronger biosignature fingerprint in emission spectra for high pressure scenarios (see 4.3.1).

*4.2.2 Nitrous oxide*

The only significant source for nitrous oxide in Earth´s atmosphere (and in our model) is the surface biotic flux. The chemical in-situ abiotic sources and sinks for nitrous oxide are several orders of magnitude smaller than the biotic source, and so they do not have a significant effect on the $N_2O$ amount in the atmosphere. The largest amount of $N_2O$ is present in the troposphere, and the main sinks of $N_2O$ are photolysis and reactions with excited oxygen ($O^1D$) (produced mostly from ozone photolysis) in the stratosphere by UV-radiation. Since $O_3$ blocks UV radiation very effectively in all scenarios, this then leads to only a small loss of nitrous oxide and, hence, only small changes in stratospheric nitrous oxide amounts (Fig 5a). Due to increasing amounts of tropospheric nitrous oxide, its spectral features are enhanced for higher pressure scenarios as seen in the emission spectra (see 4.3.2).

*4.2.3 Water*

The total amount of water (see Table 2) shows an increase for scenarios with up to 3 bar surface pressure and decreases for higher pressure scenarios (Fig. 6a). This is consistent with the

**Effect of Pressure on Earth-like planets**

increasing surface temperature and pressure. Water in the troposphere is parameterized as a function of the atmospheric temperature and relative humidity. The stratospheric production occurs mainly due to methane oxidation via hydroxyl (OH), which plays an especially important role for the lower pressure scenarios. The main processes affecting the water abundance are evaporation and condensation rather than the comparatively small chemical sources, meaning that the water fingerprint in emission spectra follows the surface temperature (see 4.3.3).

*4.2.4 Methane*

The main source for methane is a biotic source on modern Earth (and in our model) and the main destruction of methane (~95%) in the atmosphere is induced by hydroxyl. Table 2 suggests a negative feedback that suppresses the methane columns on increasing pressure. Further analysis suggests that this was not related to a response in OH which is generally a strong in-situ methane sink. Mixing processes, which determine the rate at which methane is brought from the surface to the upper levels where it is quickly oxidized, play a role. The methane amount (Fig. 7a) increases for higher surface pressure scenarios, and similar to nitrous oxide, methane shows stronger signatures in the emission spectra for higher pressure scenarios (4.3.4).

4.3 Theoretical spectral response

Here, we discuss the effect of increasing surface pressure on the spectral response of the biosignatures ozone and nitrous oxide, on the bioindicator methane, and on the climate and life-related compounds carbon dioxide and water. The spectral analysis is carried out in the infrared wavelength regime from 1 μm – 20 μm with resolutions of R=10 and R=100. Figure 8 shows the

**Effect of Pressure on Earth-like planets**

emission spectrum for the 1-bar scenario, resembling the present Earth, for a wavelength resolution of R=100. The ozone band at 9.6 μm, the nitrous oxide band at 7.8 μm, the water band at 6.3 μm, the carbon dioxide band at 14.9 μm, and the methane band at 3.3 μm are visible.

Figures 4-7 show the analysis for ozone, nitrous oxide, water, and methane respectively. Figure 9 is for carbon dioxide. Each figure consists of nine panels labeled (a) to (i). The top three panels show (a) species' concentration profiles for (b) emission spectra at R=100 and (c) brightness – black body (BB) temperature spectra at R=100. In panels (b) and (c), only the species in question is plotted as absorber. The brightness - BB temperature spectra show only the 1-bar (modern Earth), the 4-bar (highest surface temperature) and the 30-bar (highest surface pressure) scenario for better comparison.

The panels (d) to (e) show, at low and high resolution, the atmospheric absorption of the species in question, for example, $CO_2$, which contains only this absorber after subtraction of the BB planetary emissions spectrum. This was done to show more clearly the effect of increasing pressure on the absorption behavior. Panels (f) and (g), again at low and high resolution, show the emission spectra now including all absorbing species to show the effect of the absorption features in the spectrum. Within the panels (d) – (g), a zoomed-in feature shows the spectral shape in radiatively weak regimes, where in (f) - (g) the plot is shown in a logarithmic scale. For the y axis values, see panel (b).

The panels (h) - (i) show the brightness - BB temperature spectra including all absorbing species at high and low resolution respectively. Dotted lines show the spectrum when excluding the species in question. This was done to show the effect of overlapping bands. The analysis shows the effect of, for example, masking of spectral bands of minor species by dominating species. Note that this was done only to show the effect of masking features in the spectrum by different

**Effect of Pressure on Earth-like planets**

species without considering that the absence of a certain species would lead to a different atmospheric composition. Moreover, clearly the presence of other absorber species, which are not included in our scenarios, could also have masking effects.

*4.3.1 Spectral response of ozone*

Figure 4 shows the ozone features. There is absorption (panels (d) - (e)) at 4.8 μm (very weak) and 9.6 μm, which is visible at both resolutions. They become stronger at higher surface temperatures. The width of the ozone bands does not change significantly for the different scenarios. For the higher resolution (R=100) spectrum, weak bands at 5.8 μm and 14.1 μm occur.

The emission spectra (panels (f) - (g)) show that the bands at 5.8 μm and 14.1 μm are not visible, as opposed to the 9.6 μm band, which is visible in all cases. The visibility of the band at 4.8 μm decreases for higher pressure scenarios due to the occurrence of a carbon dioxide band at 4.8 μm, but is still visible at high resolution.

The brightness-BB temperature spectra (panels (h) - (i)) for the scenarios 2, 5, and 9 with (solid lines) and without (dashed lines) ozone clearly show the effect of ozone in the spectrum at 9.6 μm. When excluding ozone (dashed lines), a peak can still be seen for the 4-bar and 30-bar scenarios at 9.4 μm, which is induced by carbon dioxide. Thus, a false positive detection could be possible for atmospheres where little ozone and yet significant amounts of carbon dioxide exist. The absorption bands at 5.8 μm and 14.1 μm are masked by water (fig.6, panel (h)-(i) and carbon dioxide (fig.9, panel (h)-(i)) in the spectrum respectively. It can be seen that a peak of ozone becomes visible in the absence of water at 5.8 μm and carbon dioxide at 14.1 μm. Note that the

**Effect of Pressure on Earth-like planets**

visibility of the weak band at 5.8 μm is marginal also when excluding the masking effect of other species.

*4.3.2 Spectral response of nitrous oxide*

Figure 5 shows nitrous oxide features. The width of the nitrous oxide absorption (panels (d) - (e)), visible at 4.5 μm, 7.8 μm, and for higher pressure scenarios at 16.5 μm, increases for higher nitrous oxide amounts. The band depths are highest for the 5-bar and 10-bar scenarios at 4.5 μm and 7.8 μm, and for the 30-bar at 16.5 μm, respectively. The high resolution spectrum shows that the 7.8 μm feature is split into two bands (with a second dip at 8.5 μm). Moreover, bands at smaller wavelengths become visible. The brightness-BB spectra, considering only nitrous oxide (fig. 5c), show absorptions at 2.1 μm, 2.3 μm, 2.6 μm, 2.9 μm, 3.6 μm, 4.0 μm, and 5.3 μm, which become important for the 30-bar scenario.

The emission spectra (panels (f) - (g)) show that features of nitrous oxide are only visible at higher resolution. The feature at 8.5 μm is visible in all scenarios, and the strong band at 4.0 μm is clearly visible for the higher surface pressure scenarios. Most other bands are not visible since they are fully masked by water, methane, and carbon dioxide. The bands at 4.5 μm and 7.8 μm are visible but difficult to separate from other species feature.

This can also be seen in the brightness-BB temperature spectra (panels (h) - (i)) for the scenarios 2, 5, and 9 with (solid lines) and without (dashed lines) nitrous oxide. The visibility of nitrous oxide feature at 4.5 μm is affected by the strong carbon dioxide band at 4.3 μm, but partly visible. The 4.0 μm band offers an excellent opportunity to determine the existence of nitrous oxide, because in this region no other species (included in this work) absorb, as can be seen from

**Effect of Pressure on Earth-like planets**

comparing with the dotted lines. Also the bands at 7.8 μm- 8.5 μm are suitable to identify nitrous oxide. The peak at 8.5 μm becomes stronger for the higher pressure scenarios. The two-dip shape of this feature is affected by methane absorption. An important masking effect occurs for the strong band at 16.5 μm, where masking by carbon dioxide occurs. The reduction of carbon dioxide in the atmosphere may enable the visibility of a strong nitrous oxide feature for the three scenarios 2, 5, and 9 (see fig.9, panels (h)-(i)). Unlikely in low resolution spectra, nitrous oxide features are weakly visible at 4.0 μm only as a shoulder feature.

*4.3.3 Spectral response of water*

Figure 6 shows spectral features of water (up to 40 μm for the higher resolution). The width of the water absorption (panels (d) - (e)) increases for the higher surface temperature simulations, for which the water amount is enhanced. Absorption bands at 2.4 - 4.0 μm, ~6 μm, and at wavelengths larger than 15 μm are visible for both resolutions. Their depth increases with higher surface temperatures. Figures 6b and 6c, where water is considered as the only absorber, show three absorption bands at 1.13 μm, 1.36 μm, and 1.87 μm which become especially important for the 4-bar scenario, which has the highest surface temperature.

The emission spectra (panels (f) - (g)) show that many of the above-mentioned absorption bands can be identified at both resolutions.

The brightness-BB temperature spectra for both resolutions (panels (h) - (i)) for the scenarios 2, 5,and 9 with (solid lines) and without (dashed lines) water show that water bands mask the absorption bands of other species in the spectral regime from 1 μm - 2 μm. The bands at 2.4 μm - 4.0 μm are difficult to identify due to simultaneous absorption of carbon dioxide and methane at

**Effect of Pressure on Earth-like planets**

these wavelengths. At 2.4 μm, the slope of the spectrum begins to decrease, and there is a peak at 2.6 μm induced by water. Without water, this decrease starts at 2.65 μm induced by carbon dioxide. In the range 2.65 μm – 2.90 μm (~2.75 μm), mainly carbon dioxide determines the shape (except for the 1-bar scenario) of the spectrum at high resolution. At low resolution, these features cannot be separated. For both resolutions, at 3.07 μm, the water feature is clearly evident, and no other species considered in this work affects the spectrum. For wavelengths up to 4 μm, no more water features can be clearly identified, because, for example, at 3.3 μm a methane feature masks the water band (see fig.7, panel (h)-(i)). In the spectral range above 15 μm, water vapor features dominate the spectrum.

*4.3.4 Spectral response of methane*

Figure 7 shows methane features. Methane absorption (panels (d) - (e)) at 3.3 μm and 7.8 μm is visible and shows an increase in depth and width for higher methane amounts and pressures at both resolutions. In the high (R=100) resolution spectrum, a band at 6.5 μm is also visible. Figures 7b and 7c, where only methane is considered as absorber, suggest that for higher pressures an additional band at 2.3 μm occurs, which is best visible for the 30-bar scenario. The emission spectra (panels (f) - (g)) show that the band at 2.3 μm is visible for pressures higher than 10 bar at both resolutions. The band at 3.3 μm shows similar visibility, but can also be identified for lower pressures at the higher resolution of R=100. The band at 6.5 μm is masked by water. The methane band at 7.8 μm is hard to distinguish from the absorption of other species, such as nitrous oxide. This masking can also be seen in the brightness-BB temperature spectra (panels (h) - (i)) for the scenarios 2, 5, and 9 by comparing the spectra with (solid lines) and without (dashed lines) methane. The band at 7.8μm is strongly affected by water absorption, but also by nitrous

**Effect of Pressure on Earth-like planets**

oxide, which does not allow for a separation of the individual contributions. To distinguish these two species, the simultaneous occurrence of the methane features and of the nitrous oxide features at 7.8 μm and 8.5 μm would be an indication that nitrous oxide exists in the atmosphere and may mask a methane feature. This approach may help to infer that both species are present. For all cases, the band at 3.3 μm is visible, especially for high pressures at low and high resolution.

*4.3.5 Spectral response of carbon dioxide*

Figure 9 shows carbon dioxide features. The carbon dioxide absorption (panels (d) - (e)) at 4.3 μm and 14.9 μm are visible at low resolution, and the width of the bands increase for the higher surface pressure simulations, since the total amount of carbon dioxide also increases. The depth of the peaks is highest for the high temperature scenarios. An additional peak appears around~10 μm for the higher pressure scenarios. At the higher resolution, this band shows two peaks (at 9.4 μm and 10.5 μm). At the higher resolution, other absorption bands also appear around 4.8μm and 7.5 μm, which increase for higher pressure scenarios. Figures 9b and 9c, where only carbon dioxide is considered as absorber, show that there are also absorption bands around 1.43 μm, 1.6μm, 2.0μm, and 2.75 μm.

The emission spectra (panels (f) - (g)) show that the main bands at 4.3 μm and 14.9 μm are visible for all scenarios (where, in high resolution, the center peak may be used as an indicator of the temperature inversion in the stratosphere, although our results suggest that this feature is not detectable in the spectrum for a surface pressure of 0.5-1 bar). At lower resolution, these two bands are the only bands that are clearly discernible, since at the other wavelengths overlap with other absorption bands occurs, for example, at 2.75 μm, where the carbon dioxide absorption is

**Effect of Pressure on Earth-like planets**

not easily separated from a water feature. At the higher resolution, the bands in the lower wavelength range are visible at 1.6 μm and 2.0 μm. The bands at 4.8 μm and 10.5 μm become visible for high pressure scenarios, where the atmospheric amount of carbon dioxide increases. The bands at 1.43 μm (visible for the 30-bar scenario) and 2.75 μm are weakly visible and difficult to separate from water features. The peaks at 7.5 μm and 9.4 μm are masked by other species such as ozone.

Masking effects can be seen in the brightness-BB temperature spectra (panels (h)- (i)) for the scenarios 2, 5, and 9 by comparing the spectra with (solid lines) and without (dashed lines) carbon dioxide. At 4.3 μm, it can be seen that the absorption band of nitrous oxide at 4.5 μm is affected by carbon dioxide. At the lower resolution, it can be seen that the carbon dioxide bands at 1.6 μm and 2.0 μm are visible for the 30-bar scenario. This region is, however, partly masked by water features. Moreover, at higher resolution the band at 9.4 μm is masked by ozone for all cases, while the band at 7.5 μm is masked by methane and nitrous oxide. An important masking also occurs of the band around 14.9 μm, where water features at wavelengths larger than 11 μm are masked. The dashed lines show that carbon dioxide affects these parts of the spectrum strongly. In the absence of carbon dioxide, a nitrous oxide feature would dominate at 16.5μm for all scenarios, and also a weak ozone feature at 14.1 μm becomes visible.

Figure 10 shows the effect of varying the carbon dioxide amount upon the absorption band around 16.5 μm in the higher resolution brightness-BB temperature spectrum for the 30 bar surface pressure scenario. A strong nitrous oxide absorption shape becomes visible for low carbon dioxide amounts. Note that, in Figure 10, we only show the effect of varying $CO_2$ concentration upon the band strengths and ignore any change in temperature and other atmospheric species profiles in order to provide a first estimate of how much carbon dioxide is

**Effect of Pressure on Earth-like planets**

required to conceal the $N_2O$ band. More detailed estimates would require consistent climate-chemistry calculations.

## 5. Discussion

This work investigates the effect of different atmospheric mass on the climate, chemistry, and spectral appearance of Earth-like planets. An important aspect in this work is the opposing effects of warming by increasing greenhouse gases amounts and the cooling by Rayleigh scattering, which determine the resulting surface temperature. Uncertainties that affect the temperature, for example, thermal radiative fluxes and numerical limitations of the model are discussed below.

A numerical limitation in the model that affects the surface temperature is the implementation of the convective surface layer. The lowest model layer is always set to be convective. For the high pressure scenarios with colder surface temperatures, this layer is possibly not convective, and thus the model would overestimate the surface temperature. Surface albedo can have a potentially strong influence upon the surface temperature simulated in the model. The surface albedo is tuned to our modern day Earth run to reproduce the surface temperature of present-day Earth in order to consider the effect of clouds in a straightforward manner. Dense atmospheres could also favor enhanced cloud and haze formation (consider, e.g., Venus). Nitrous oxide, a strong greenhouse gas, is not considered in the thermal radiative transfer, and thus the atmospheric temperature could be underestimated. To validate the results of the 1D model, the upwards propagating radiative flux calculated by the 1D model are compared to those of the line-by-line (lbl) model SQuIRRL (Schreier & Böttger (2003)).

**Effect of Pressure on Earth-like planets**

Moderate deviations in the stratosphere and most of the troposphere were found by comparing the upwards flux, which is a common issue in climatic simulations (see Myhre et al. (1998), Iacono et al. (2008)). Figure 11a shows the ratio of the upwards flux calculated by the atmospheric column model divided by the upwards flux calculated by the lbl-model. The red solid lines show the agreement to within 95% and the green solid lines within 90%. Beginning with 0.5 bar surface pressure, the stratospheric agreement in upwards flux between both model calculations decreases for the higher surface pressure scenarios and is present in each simulation. Up to 10 bar surface pressure, the agreement in the stratosphere is more than 95 %, for higher pressure scenarios more than 92 %. For the troposphere and near surface, the agreement decreases to ~87 % for the 5-bar scenario and increases afterwards for the other scenarios within a range of 90% to 95%. Overall, the upwards radiative flux calculated with the thermal transfer scheme is underestimated in the stratosphere and overestimated in the troposphere compared to the lbl- calculations. In comparison with the 1bar modern Earth validated simulation, these values are acceptable and comparable to agreements in climatic simulations as shown, e.g., in the works of Goldblatt et al. (2009) and Zsom et al. (2013).

Different models produce different radiative responses due to, for example, the spectral range employed, molecular interactions, and the chemical species considered. Figure 11b compares results from this work (using the thermal transfer scheme MRAC Complete (Modified RRTM for Application in $CO_2$-dominated atmospheres)) compared with results of similar simulations performed by using a thermal transfer scheme, which is valid over a wider pressure range up to 1000 bar called MRAC Simple (Philip von Paris, PhD Thesis (2010)). This scheme, however, only considers absorption of water and carbon dioxide. Similar to the previously presented simulations, the surface temperature increases with increasing surface pressure here (up to 3 bar) and decreases afterwards for the other scenarios. The surface temperature in both models deviate

**Effect of Pressure on Earth-like planets**

for each scenario on the order of a few Kelvin. The cooling of the atmosphere by the scattering effect for high pressure atmospheres is confirmed. The deviation in surface temperatures between the model results using the two different thermal transfer schemes can be explained by the missing contributions of methane and ozone. This, together with the moderate deviation from the lbl radiative fluxes, suggests that the conclusions of this work should be robust.

**6. Summary & Conclusion**

A cloud-free 1D global photochemical climate model was used to simulate Earth-like ($N_2$-$O_2$ dominated) atmospheres for Earth-sized planets with different atmospheric surface pressures between 0.5 bar and 30 bar orbiting at 1 AU around the Sun. The surface temperatures change significantly for the different surface pressure scenarios. From the 0.5 bar to 4 bar surface pressure scenario, the surface temperature increases from 285.3 K up to 295.0 K and then decreases to 274.8 K for the 30 bar surface pressure scenario. The increase up to 4 bar surface pressure is induced by the increasing greenhouse effect. For further increases in surface pressure, the cooling induced by back-scattering of the incoming shortwave radiation to space due to Rayleigh scattering is stronger than the greenhouse heating and the surface temperature decreases. A strong change in the species concentration occurs for water. Larger surface temperatures lead to an increase in water amounts from 0.5 bar to4 bar surface pressure. The lower temperatures for the scenarios with higher surface pressure lead to a decrease in the water content. The ozone amount, which is important for the UV-radiation protection of the surface, did not change greatly due to the ozone-UV-feedback. Smog ozone in the lower atmosphere increased for higher surface pressure scenarios, which could be harmful for life. The nitrous oxide and methane amounts increased for higher pressure scenarios. In the second part of this

**Effect of Pressure on Earth-like planets**

work, we investigated the effect of the different pressure scenarios on spectral response. For this, theoretical emission- and brightness temperature spectra were calculated at resolutions of R~10 (low) and R~100 (high). All in all, the spectral features showed an increase in band width for higher surface pressure.

In the lower resolution emission spectra, water and carbon dioxide were visible for all scenarios. Their strong absorptions can mask the absorption bands of other species, for example, nitrous oxide. Ozone can be identified partly as a shoulder feature. Methane can be estimated for lower pressure scenarios and is visible for scenarios with surface pressure higher than 10 bar, but the absorption occurs in a regime where only little radiation is emitted. Nitrous oxide is not visible.

In the higher resolution emission spectrum, all analyzed species are visible for all pressure scenarios. The analysis of the brightness temperature spectra showed that the absorption of the species depends in some cases on their higher amounts, which usually increase with pressure and, in other cases, on higher surface temperature.

*Ozone:* The biosignature ozone showed an increase in the band depth at 9.6 μm for the higher temperature and pressure scenarios in comparison to the modern Earth scenario. At 5.8 μm and 14.1 μm, ozone bands also appear that are masked by water and carbon dioxide, respectively. The 4.8 μm band becomes more important for carbon dioxide-poor atmospheres and thus modern Earth and the 0.5-bar scenario, but nevertheless it is difficult to distinguish from a carbon dioxide feature.

*Nitrous oxide:* Carbon dioxide also overlaps with the nitrous oxide band at 4.5 μm, which makes separating the two difficult, but possible, at high resolution. Another masking effect that can be important occurs at 16.5 μm; here, another strong nitrous oxide feature is visible, but is again

**Effect of Pressure on Earth-like planets**

masked by carbon dioxide. In a low carbon dioxide atmosphere, this feature could possibly be an important indication for life. Nitrous oxide features are only visible at high resolution at 4.0 μm and 7.8-8.5 μm and become stronger for scenarios with higher pressure and thus with higher amounts of $N_2O$. There is a characteristic shape that shows two peaks (with a local maximum in the band center) of nitrous oxide in some bands, for example 7.8 - 8.5 μm. The first peak at 7.8 μm is masked by methane, but the second one at 8.5 μm is clearly visible.

*Methane:* The strong band of the bioindicator methane at 7.8 μm interacts strongly with a nitrous oxide (and also water) band, which leads to a broadening of this band for higher pressure scenarios. Nevertheless, the methane feature at 3.3 μm is visible for different scenarios. The band at 2.3 μm indicates a high methane atmosphere and becomes visible for high methane amounts.

*Carbon dioxide:* As mentioned, the life-related compounds water and carbon dioxide mask different features due to strong absorption. Due to higher carbon dioxide amounts for higher pressure scenarios various bands appear that are absent in the modern Earth-scenario, for example, at 1.6 μm or 4.8 μm, which become strong for higher pressures. The depth of the bands at 4.3 μm and 14.9 μm depends on the temperature rather than on the surface pressure. Other bands, for example at 1.6 μm and 2.0 μm, increase for higher carbon dioxide amounts. Similar to nitrous oxide, carbon dioxide shows a local maximum in the band center. The center of the 4.3 μm and 14.9 μm band is an indication for the temperature inversion at the stratosphere. This peak increases for higher pressure scenarios, where the temperature inversion also increases. This peak disappears for the low pressure scenarios, although a temperature inversion takes place.

*Water:* Another life-related compound with strong absorption bands is water. The band depths of water are strongest for the 3-bar and 4-bar scenarios where water reaches the highest amount in the atmosphere. The strong water bands are visible at different wavelengths, for example 1 μm -2

**Effect of Pressure on Earth-like planets**

μm, and the strongest band at ~6 μm. The bands at wavelengths >15 μm affect the whole spectrum beyond 20 μm.

In conclusion, in some of the cases considered water and carbon dioxide have a strong masking effect in the spectra, which influences the visibility of other species, for example, the biosignatures ozone and nitrous oxide. Nevertheless, such signatures, in general, are enhanced and could be easier to identify for an Earth-like planet with higher surface pressure compared with that of modern Earth.

# Effect of Pressure on Earth-like planets

**Effect of Pressure on Earth-like planets**

**Effect of Pressure on Earth-like planets**

**Effect of Pressure on Earth-like planets**

**Effect of Pressure on Earth-like planets**

**Tables and Figures:**

Table 1: Scenarios in this work

(✓ = included ; ✘ = not included)

| Scenario | $P_0$ [bar] | Chemistry/Climate profiles & Emission spectra | Brightness temperature spectra |
|---|---|---|---|
| 1 | 0.5 | ✓ | ✘ |
| 2 | 1 | ✓ | ✓ |
| 3 | 2 | ✓ | ✘ |
| 4 | 3 | ✓ | ✘ |
| 5 | 4 | ✓ | ✓ |
| 6 | 5 | ✓ | ✘ |
| 7 | 10 | ✓ | ✘ |
| 8 | 20 | ✓ | ✘ |
| 9 | 30 | ✓ | ✓ |

Table 2: Surface temperature and atmospheric species column densities for each pressure scenario simulated with the 1D coupled photochemical climate model (considering Rayleigh scattering)

| $P_0$ [bar] | $T_{Surf}$ [K] | $O_3$ | $H_2O$ | $N_2O$ | $CH_4$ |
|---|---|---|---|---|---|
| **0.5** | 283.25 | 274.4 (0.90) | $2.09 \times 10^5$ (0.82) | 115.6 (0.49) | 966 (0.80) |
| **1** | 288.26 | 305.1 (1) | $2.56 \times 10^5$ (1) | 233.8 (1) | 1207 (1) |
| **2** | 292.45 | 315.6 (1.03) | $2.93 \times 10^5$ (1.15) | 465.8 (1.99) | 1666 (1.38) |
| **3** | 294.25 | 329.2 (1.08) | $3.05 \times 10^5$ (1.19) | 695.9 (2.97) | 2270 (1.88) |
| **4** | 294.97 | 343.3 (1.12) | $3.03 \times 10^5$ (1.18) | 925.5 (3.96) | 3014 (2.49) |
| **5** | 294.58 | 356.7 (1.17) | $2.88 \times 10^5$ (1.13) | 1154.7 (4.94) | 3743 (3.10) |
| **10** | 289.18 | 409.4 (1.34) | $2.29 \times 10^5$ (0.89) | 2298.8 (9.83) | 7100 (5.88) |
| **20** | 280.53 | 477.0 (1.56) | $1.37 \times 10^5$ (0.54) | 4585.3 (19.61) | 16780 (13.90) |
| **30** | 274.75 | 509.4 (1.67) | $0.91 \times 10^5$ (0.35) | 6860.0 (29.34) | 29191 (24.18) |

3333

Table 3: Considered species and included effects in the model for the solar- and thermal radiative transfer scheme

| Species | Solar | Thermal | Continuum | Rayleigh scattering | Heat Capacity | Lapse Rate |
|---|---|---|---|---|---|---|
| $N_2$ | ✗ | ✓ | ✗ | ✓ | ✓ | ✗ |
| $O_2$ | ✓ | ✗ | ✗ | ✓ | ✓ | ✗ |
| Ar | ✗ | ✗ | ✗ | ✗ | ✓ | ✗ |
| $H_2O$ | ✓ | ✓ | ✓ | ✓ | ✓ | ✓ |
| $CO_2$ | ✓ | ✓ | ✓ | ✓ | ✓ | ✓ |
| CO | ✗ | ✗ | ✗ | ✓ | ✓ | ✗ |
| $CH_4$ | ✓ | ✓ | ✗ | ✓ | ✓ | ✗ |
| $O_3$ | ✓ | ✓ | ✗ | ✗ | ✗ | ✗ |
| $H_2$ | ✗ | ✗ | ✗ | ✓ | ✓ | ✗ |
| He | ✗ | ✗ | ✗ | ✓ | ✓ | ✗ |
| CO | ✗ | ✗ | ✗ | ✓ | ✓ | ✗ |

Table 4: Initial start and boundary parameter used in the climate & chemistry model

| Initial and Boundary Parameters | | Earth tuned outgassing, emission & deposition | |
|---|---|---|---|
| $N_2$ | ~ 0.78 | | |
| $O_2$ | 0.21 | Emission in molecules/s*cm$^2$: | |
| $CH_4$ | $1.6*10^{-6}$ | | |
| Ar | 1 % | | |
| $CO_2$ | $3.55*10^{-4}$ | $CH_4$ | $1.24*10^{11}$ |
| He | $1*10^{-8}$ | CO | $2.39*10^{11}$ |
| Relative humidity (Manabe & Wetherald (1967)) | 77% | $N_2O$ | $1.15*10^9$ |
| Distance to Sun | 1 AU | $CH_3Cl$ | $2.52*10^8$ |
| Star | Sun | | |
| Planetary radius: | 6378 km | Effusion in molecules/s*cm$^2$: | |
| P (TOA) | $6.6 * 10^{-5}$ bar | CO | $-5.07*10^8$ |
| P (Surface) | 1 bar | O | $-5.07*10^8$ |
| Zenith angle: Chemistry module Climate module | 45° 60° | Volcanic outgassing in Tg/yr : | |
| Gravity: | 9.81 m/s$^2$ | $SO_2$ | 60 |
| Surface albedo | 0.2155 | $H_2S$ | 3 |

**Effect of Pressure on Earth-like planets**

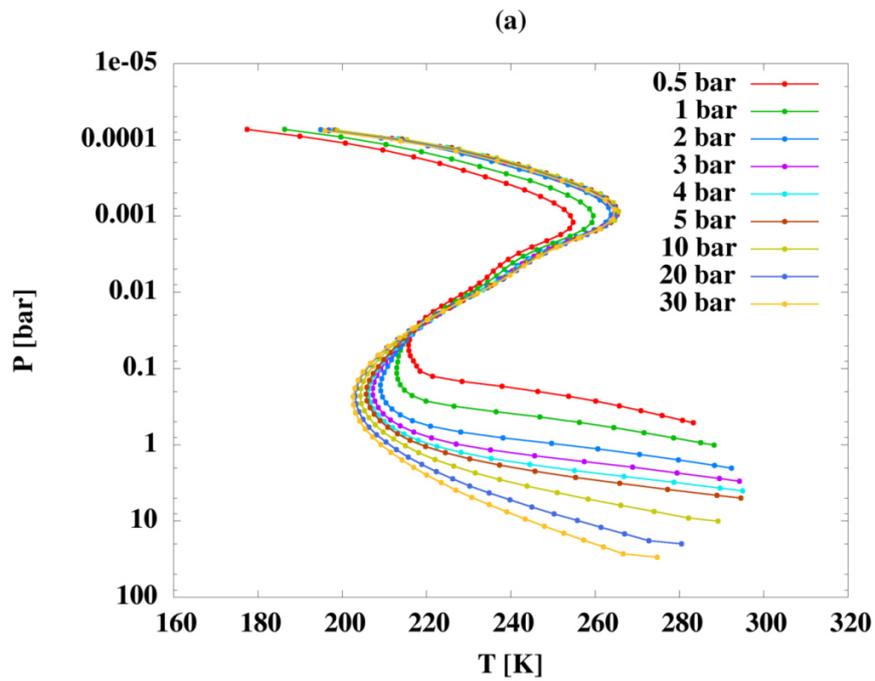

**FIG.1(a).** Pressure-temperature profiles for different surface pressures

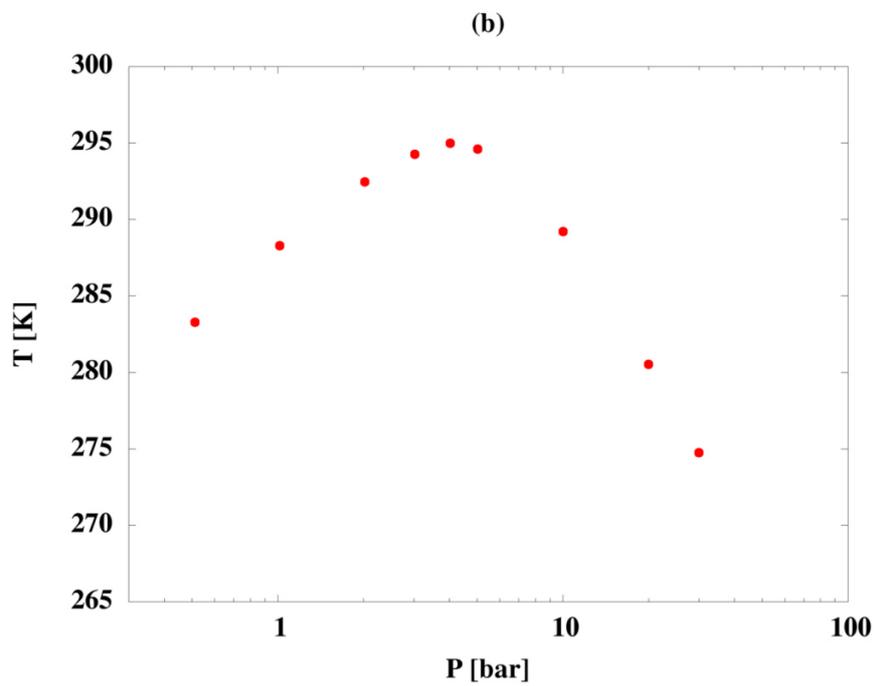

**FIG.1(b).** Different pressure Earth-like atmosphere simulations: Surface temperature for different surface pressure scenarios

**Effect of Pressure on Earth-like planets**

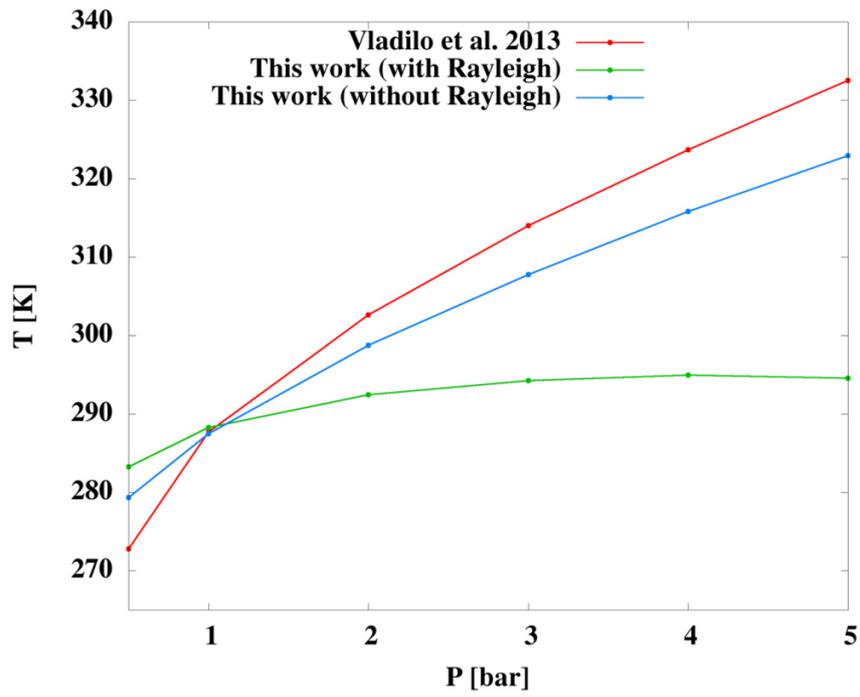

**FIG.2.** Surface temperature for different surface pressure scenarios

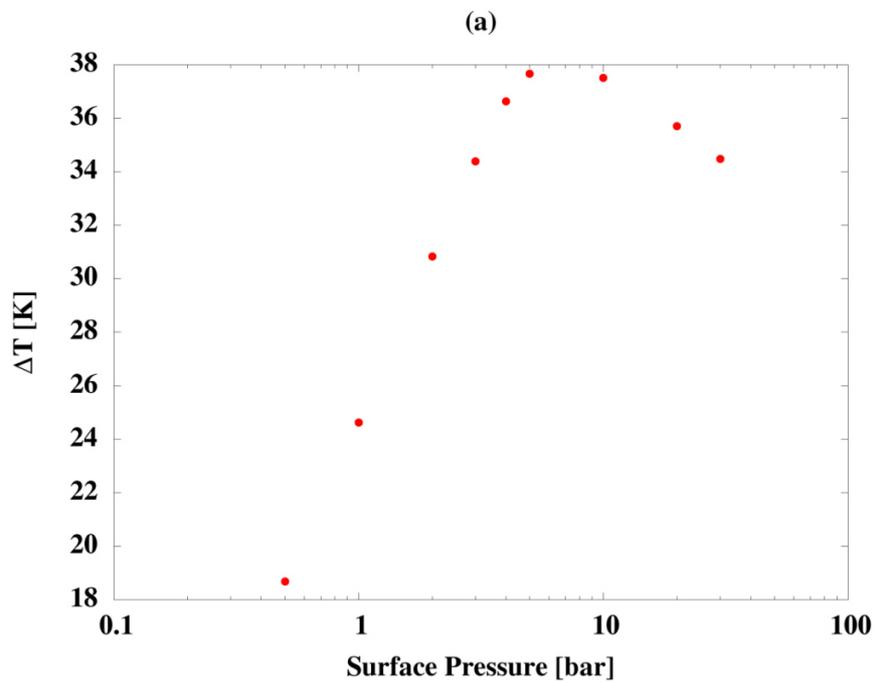

**FIG.3(a).** Greenhouse effect for different surface pressure scenarios

**Effect of Pressure on Earth-like planets**

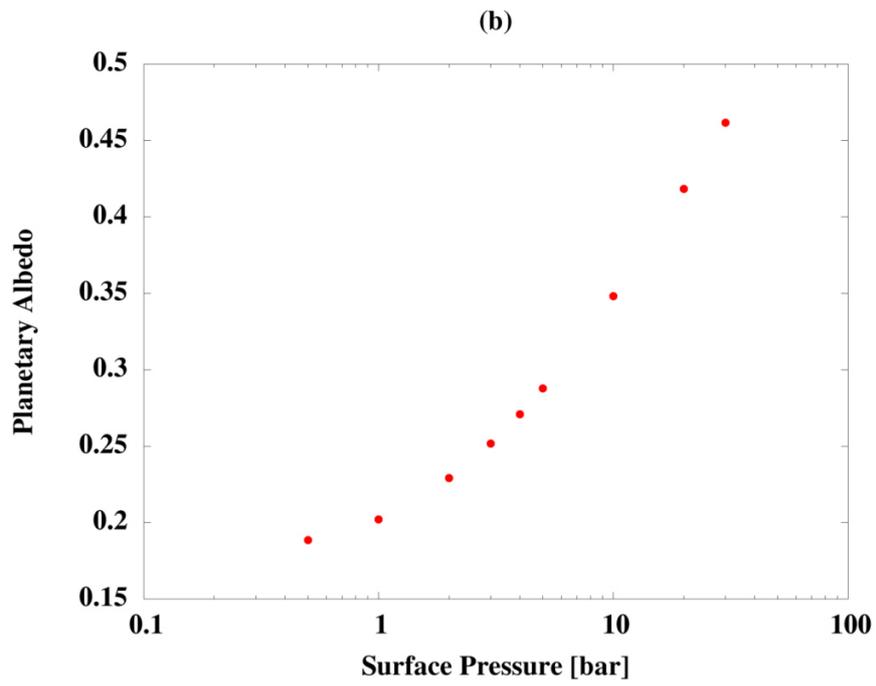

**FIG. 3(b).** Planetary albedo for different surface pressure scenarios

**Effect of Pressure on Earth-like planets**

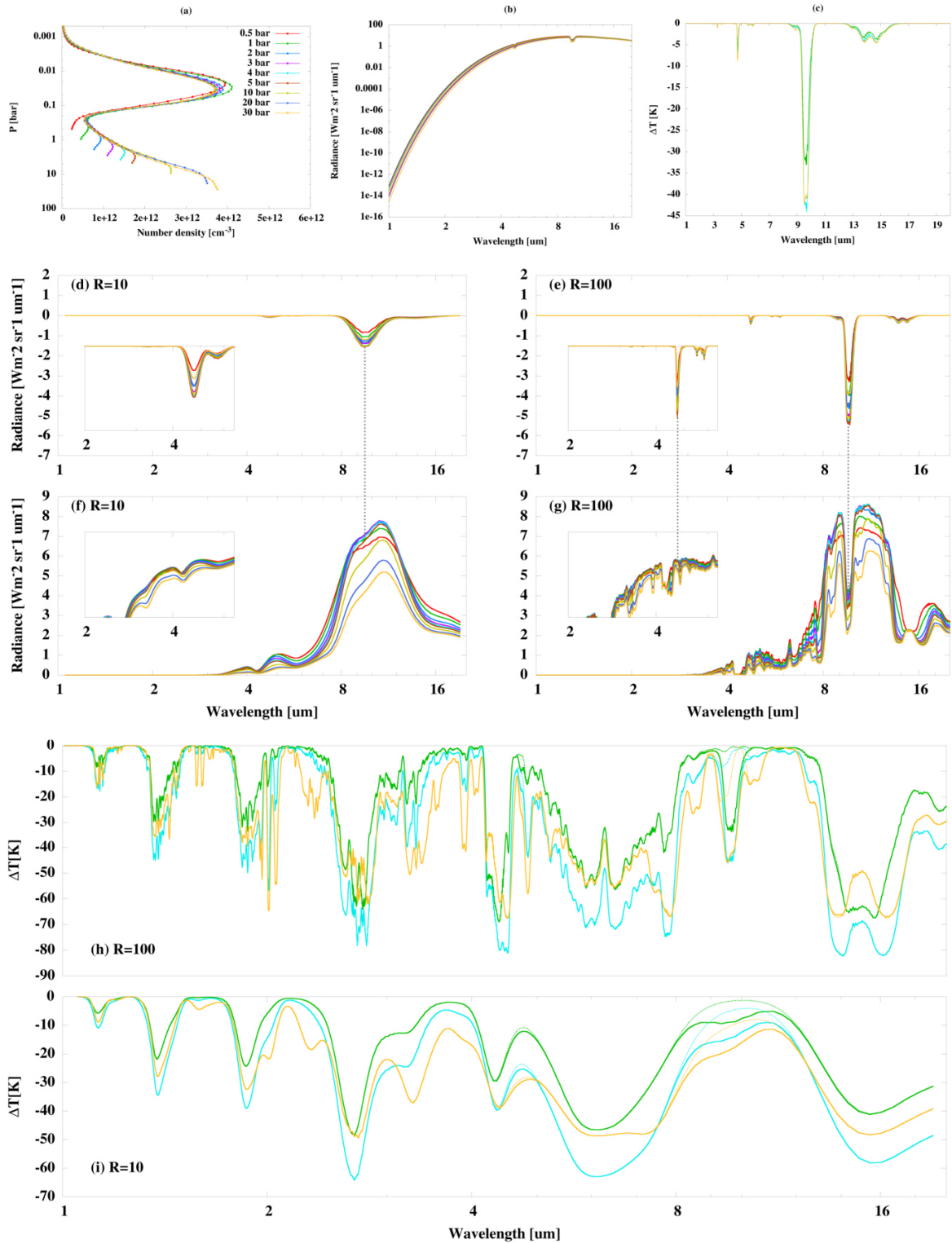

**FIG.4.** $O_3$- profile (a), Considering only ozone as absorber planetary emission- (b) and B-T- spectrum (subtracted from a BBODY spectrum) (c), low & high resolution absorption feature spectra ((d) - (e)), as well

**Effect of Pressure on Earth-like planets**

for the diff. scenarios emission spectra ((f) - (g)) and B-T-spectra (subtracted from a BBODY spectrum) ((h) - (i)) with dashed lines indicating the absence of ozone. Grey dotted lines help to indicate spectral features. Within the panels ((d) - (g)) small windows show a zoomed-in panel for the radiative weak regimes, where in ((f) - (g)) the plot is on a logarithmic scale. The tic-marks for the logarithmic scale are shown in steps of 1 micron from 1 to 10 microns and in steps of 2 microns above 10 microns. For the y axis values, see panel (b).

**Effect of Pressure on Earth-like planets**

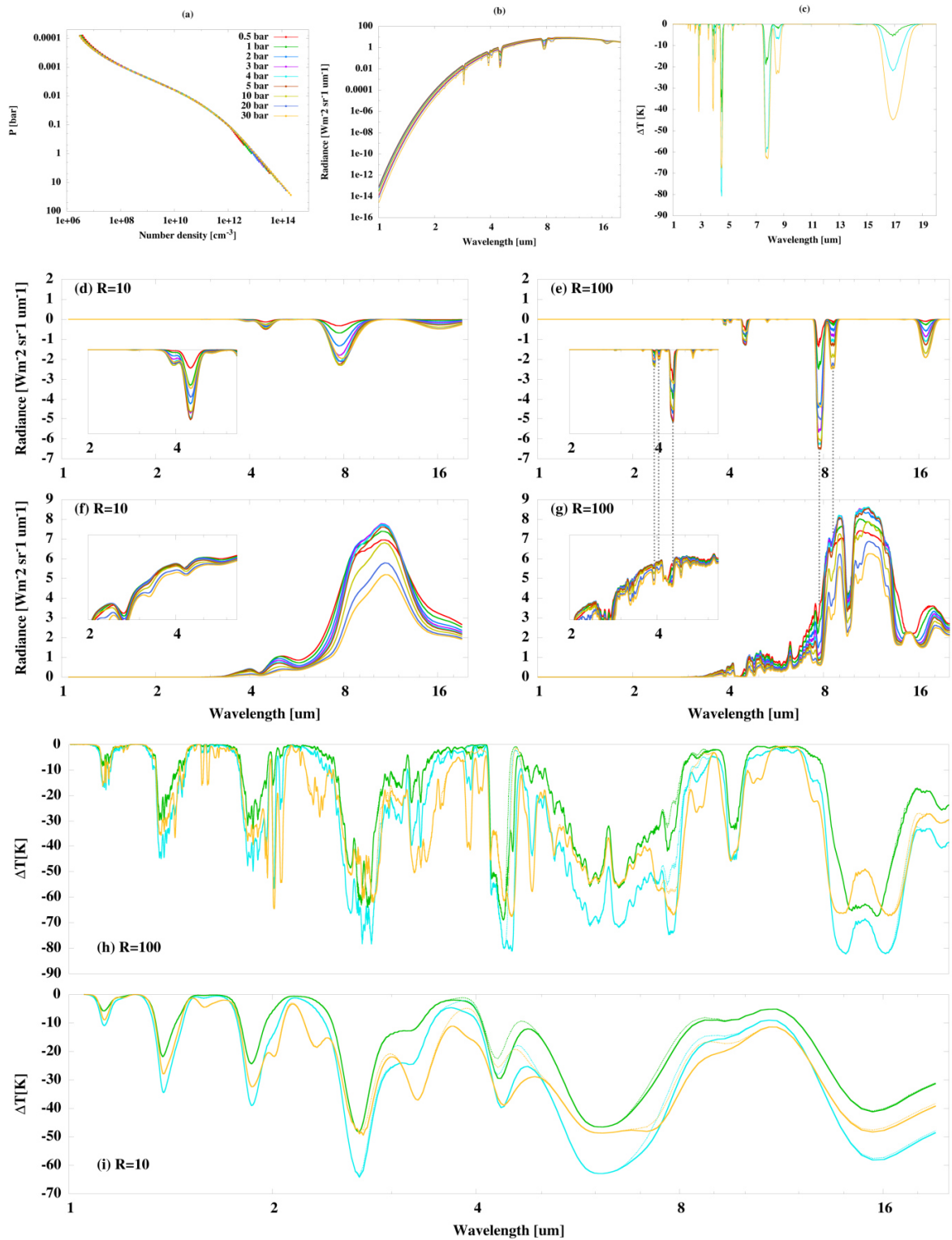

**FIG.5.** N$_2$O- profile (a), Considering only N$_2$O as absorber planetary emission- (b) and B-T- spectrum (subtracted from a BBODY spectrum) (c), low & high resolution absorption feature spectra ((d) - (e)), as well

**Effect of Pressure on Earth-like planets**
for the diff. scenarios emission spectra ((f) - (g)) and B-T-spectra (subtracted from a BBODY spectrum) ((h) - (i)) with dashed lines indicating the absence of $N_2O$. Grey dotted lines help to indicate spectral features. Within the panels ((d) - (g)) small windows show a zoomed-in panel for the radiative weak regimes, where in ((f) - (g)) the plot is on a logarithmic scale. The tic-marks for the logarithmic scale are shown in steps of 1 micron from 1 to 10 microns and in steps of 2 microns above 10 microns. For the y axis values, see panel (b).

## Effect of Pressure on Earth-like planets

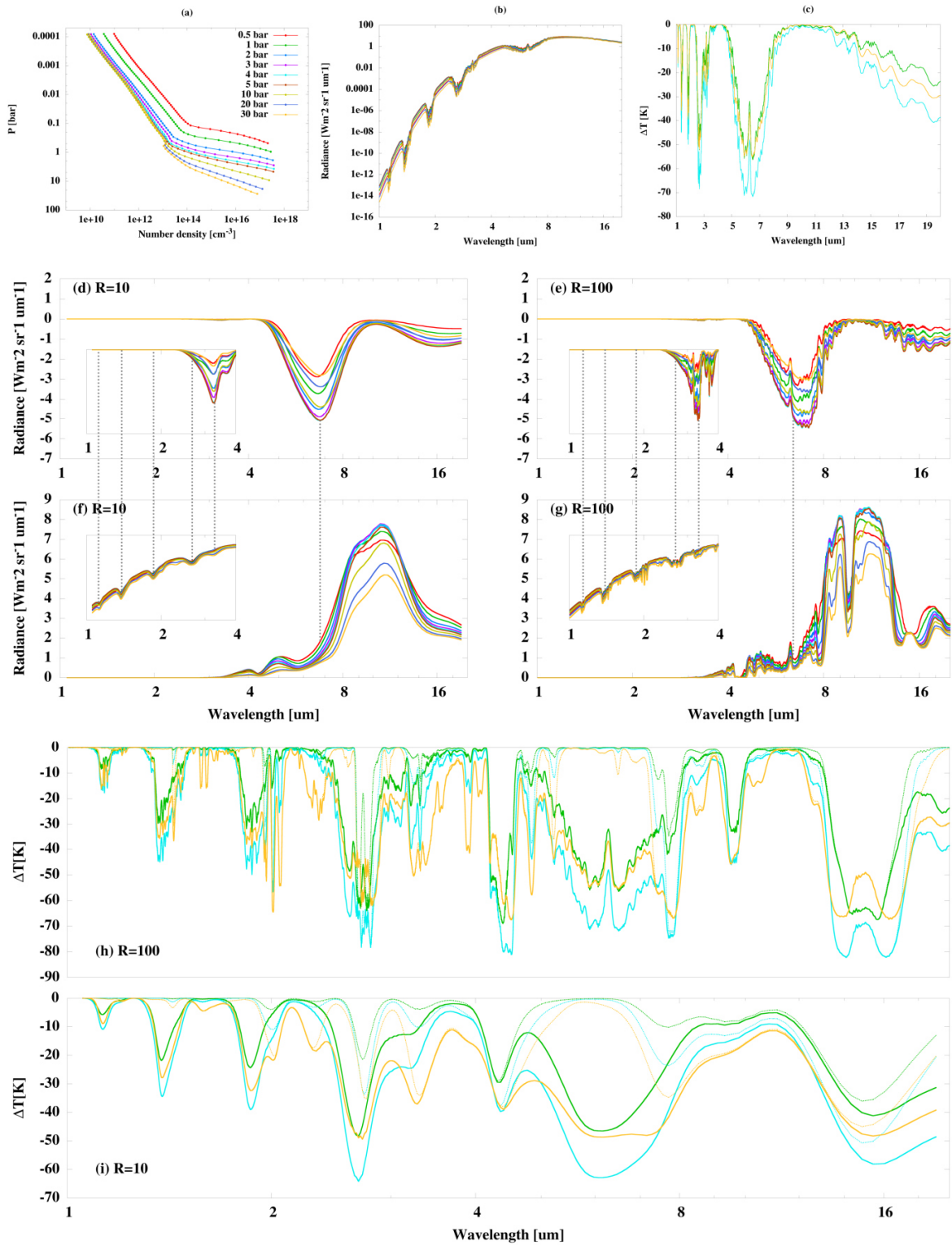

**FIG.6.** $H_2O$- profile (a), Considering only water as absorber planetary emission- (b) and B-T- spectrum (subtracted from a BBODY spectrum) (c), low & high resolution absorption feature spectra ((d) - (e)), as well

**Effect of Pressure on Earth-like planets**

for the diff. scenarios emission spectra ((f) - (g)) and B-T-spectra (subtracted from a BBODY spectrum) ((h) - (i)) with dashed lines indicating the absence of water. Grey dotted lines help to indicate spectral features. Within the panels ((d) - (g)) small windows show a zoomed-in panel for the radiative weak regimes, where in ((f) - (g)) the plot is on a logarithmic scale. The tic-marks for the logarithmic scale are shown in steps of 1 micron from 1 to 10 microns and in steps of 2 microns above 10 microns. For the y axis values, see panel (b).

# Effect of Pressure on Earth-like planets

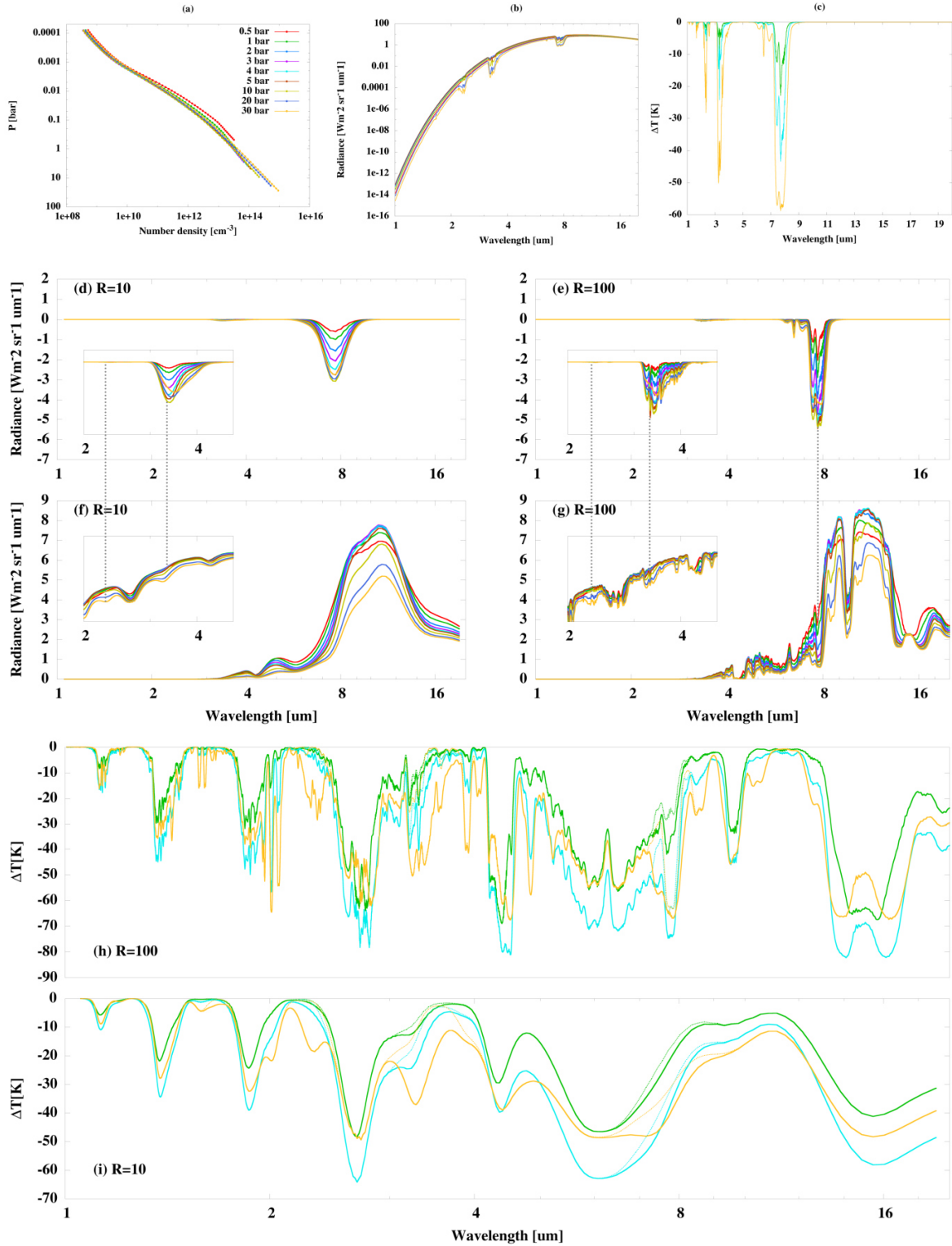

**FIG.7.** $CH_4$- profile (a), Considering only methane as absorber planetary emission- (b) and B-T- spectrum (subtracted from a BBODY spectrum) (c), low & high resolution absorption feature spectra ((d) - (e)), as well

**Effect of Pressure on Earth-like planets**

for the diff. scenarios emission spectra ((f) - (g)) and B-T-spectra (subtracted from a BBODY spectrum) ((h) - (i)) with dashed lines indicating the absence of methane. Grey dotted lines help to indicate spectral features. Within the panels ((d) - (g)) small windows show a zoomed-in panel for the radiative weak regimes, where in ((f) - (g)) the plot is on a logarithmic scale. The tic-marks for the logarithmic scale are shown in steps of 1 micron from 1 to 10 microns and in steps of 2 microns above 10 microns. For the y axis values, see panel (b).

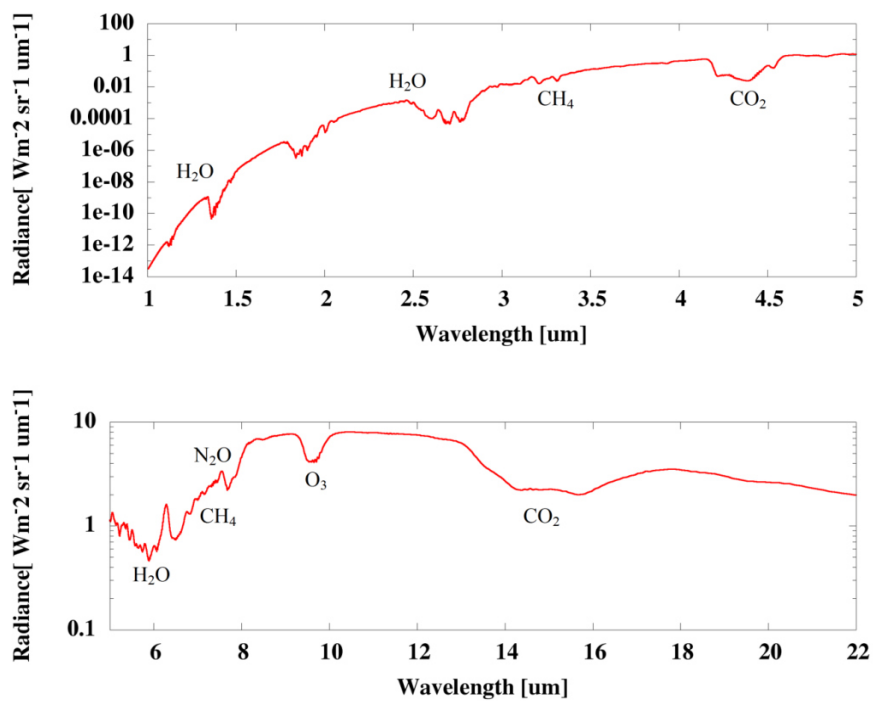

**FIG.8.** Biosignature Emission Spectrum for present Earth simulation with 1 bar surface pressure

**Effect of Pressure on Earth-like planets**

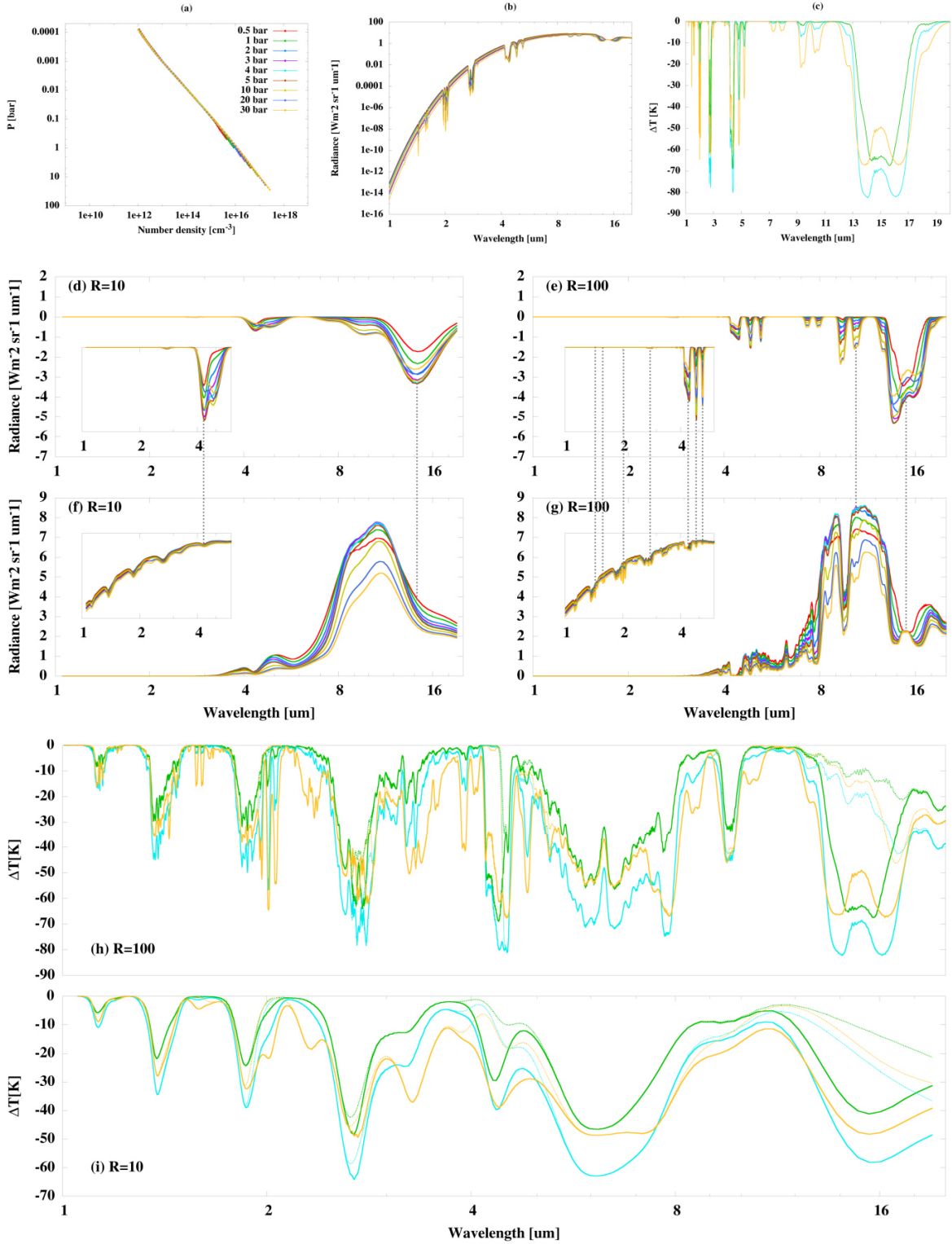

**FIG.9.** $CO_2$- profile (a), Considering only carbon dioxide as absorber planetary emission- (b) and B-T-spectrum (subtracted from a BBODY spectrum) (c), low & high resolution absorption feature spectra ((d) - (e)),

**Effect of Pressure on Earth-like planets**

as well for the diff. scenarios emission spectra ((f) - (g)) and B-T-spectra (subtracted from a BBODY spectrum) ((h)-(i)) with dashed lines indicating the absence of $CO_2$. Grey dotted lines help to indicate spectral features. Within the panels ((d) - (g)) small windows show a zoomed-in panel for the radiative weak regimes, where in ((f) - (g)) the plot is on a logarithmic scale. The tic-marks for the logarithmic scale are shown in steps of 1 micron from 1 to 10 microns and in steps of 2 microns above 10 microns. For the y axis values, see panel (b).

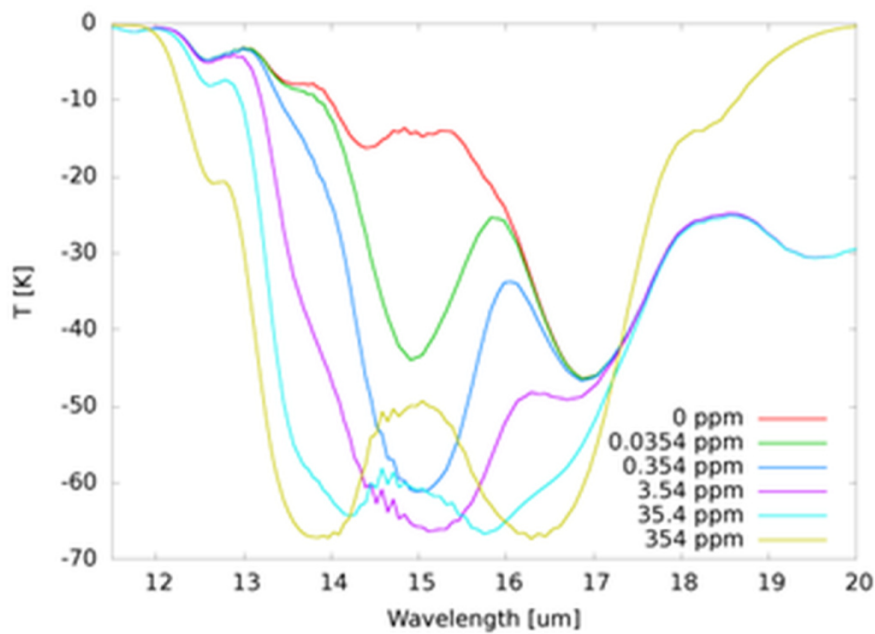

**FIG.10.** Brightness Temperature- Spectrum for the 30bar scenario with different levels of carbon dioxide (R=100)

**Effect of Pressure on Earth-like planets**

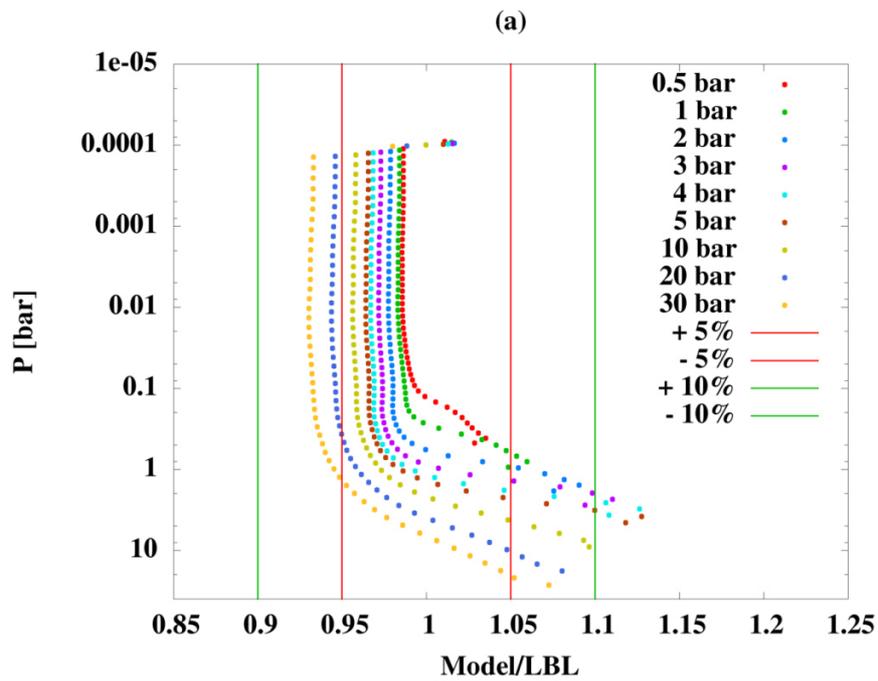

**FIG.11(a).** Upwards flux-profiles: Comparison between MRAC Complete and SQuIRRL

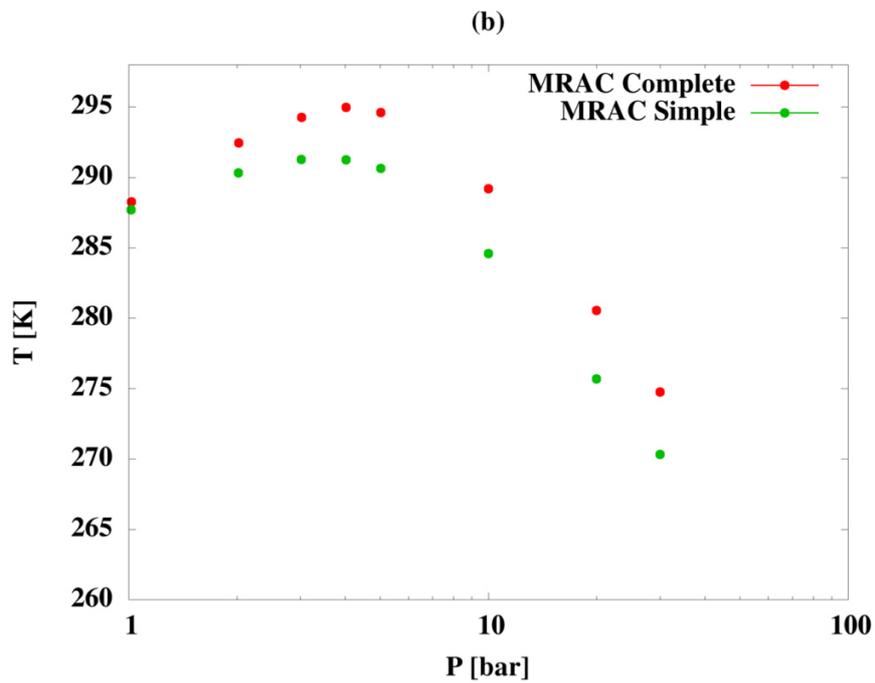

**FIG.11(b).** High pressure Earth-like atmosphere simulations: Surface temperature for different surface pressure simulations